\documentclass[usenatbib]{mnras}

\usepackage{newtxtext,newtxmath}
\usepackage[T1]{fontenc}

\DeclareRobustCommand{\VAN}[3]{#2}
\let\VANthebibliography\thebibliography
\def\thebibliography{\DeclareRobustCommand{\VAN}[3]{##3}\VANthebibliography}

\usepackage{graphicx}	
\graphicspath{{Images/}}
\usepackage{amsmath}	
\usepackage[table,xcdraw]{xcolor}    
\usepackage{subfig}
\usepackage{comment}
\usepackage[normalem]{ulem}
\usepackage{physics}
\usepackage{xcolor}
\usepackage{lipsum}
\usepackage{xfrac}
\usepackage[labelfont=bf]{caption}

\definecolor{royalazure}{rgb}{0.0, 0.22, 0.66}
\definecolor{darkgreen}{rgb}{0.06, 0.46, 0.06}
\definecolor{auburn}{rgb}{0.43, 0.21, 0.1}
\definecolor{bostonuniversityred}{rgb}{0.8, 0.0, 0.0}

\usepackage{hyperref}
\hypersetup{
    colorlinks=true,
    linkcolor=bostonuniversityred,
    linktoc=all,
    filecolor=blue,      
    urlcolor=royalazure,
    citecolor=royalazure
}

\title[Planet Formation in a Warped Disc]{Planet Formation in a Warped Disc: Gravitational Instability and Misaligned Infall}

\author[S. Rowther et al.]{Sahl Rowther,$^{1,2,3}$\thanks{E-mail: sahl.rowther.1@warwick.ac.uk},
Josh Calcino$^{4}$ and Richard Alexander$^3$
\\
$^{1}$Centre for Exoplanets and Habitability, University of Warwick, Coventry CV4 7AL, UK\\
$^{2}$Department of Physics, University of Warwick, Coventry CV4 7AL, UK\\
$^{3}$School of Physics and Astronomy, University of Leicester, Leicester LE1 7RH, UK\\
$^{4}$Department of Astronomy, Tsinghua University, 30 Shuangqing Rd, 100084 Beijing, China
}

\date{Accepted XXX. Received YYY; in original form ZZZ}

\pubyear{2026}

\begin{document}
\label{firstpage}
\pagerange{\pageref{firstpage}--\pageref{lastpage}}
\maketitle 

\begin{abstract}
We perform three-dimensional smoothed particle hydrodynamics simulations to investigate the impact of misaligned infall on a gravitationally unstable protoplanetary disc. The interaction between the disc and the infalling gas does not occur uniformly across all radii. As infall impacts the outer disc first, the outer disc experiences stronger torques than the inner disc, resulting in a warped disc where the angular momentum vector of the disc varies with radius. The surrounding infalling gas reprocesses stellar irradiation and re-emits it back towards the disc, resulting in a warmer disc. The higher temperatures cause the disc to become more gravitationally stable resulting in weaker spiral arms. However, over time, infall also continuously increases the disc mass, gradually altering the local conditions and strengthening gravitational instabilities despite the azimuthally averaged Toomre $Q$ indicating that the disc is stable against gravitational instability. Eventually, one of the spiral arms fragments, forming a giant planet by gravitational instability in a disc warped by misaligned infall. Hence, the impact of infall on gravitational instability in general will vary depending on the infall and disc properties due to the balance between the stabilising and destabilising effects. We also find that infall alters the expected signatures of gravitationally unstable discs by introducing asymmetries in scattered-light, continuum, and molecular line emission.
\end{abstract}




\section{Introduction}

Observations in scattered-light \citep{2024Garufi} and molecular line emission \citep{1987Sargent,2020Pineda,2024Valdivia-Mena,2026DeSimone} have revealed that protoplanetary discs do not evolve isolated from the surrounding star-forming environment. Scattered-light observations extending further out to the kilo-au scale have revealed streamer-like structure \citep{2023Gupta}, and ambient material from the environment in 15\% of the discs in \citet{2024Garufi}. Observations of molecular line emission have also revealed streamer candidates in 7 out of the 16 discs in \citet{2024Valdivia-Mena}. 

Simulations of isolated, coplanar discs viewed edge-on predict symmetric scattered-light emission from the upper and lower disc surfaces separated by a dark lane associated with the optically thick midplane. However, some observations of edge-on discs reveal more complex features in the form of asymmetric tail-like structures as in HH 30 \citep{2025Tazaki} and lateral asymmetries seen in 15 out of the 20 discs in \citet{2024Villenave}.

The lateral asymmetries can be explained by a warped disc, where the upper \& lower disc surfaces are illuminated differently, breaking the symmetrical emission expected from coplanar discs \citep{2019Nealon,2024Villenave,2025Kimmig}. A disc is considered warped when its angular momentum vector varies with radius. Warping a disc requires a differential torque such that the rate of change of the local angular momentum vector of the disc depends on radius causing each annulus to precess at different rates. There are multiple scenarios that are able to warp a disc.  These include a misaligned external companion \citep{2018Zanazzi,2020Gonzalez,2020bNealon}, a disc around a misaligned binary \citep{2013Facchini,2020Nealon,2023Young,2025Rowther}, a misaligned massive planet \citep{2018Nealon,2019Zhu}, or a stellar flyby inclined relative to the disc \citep{2016Xiang,2019Cuello,2020aNealon}. 

Another scenario is through infall from the surrounding star-forming environment \citep{2021Kuffmeier}. If the angular momentum of the infalling gas is misaligned relative to the disc, it exerts a differential torque that warps the disc. Simulations that model disc formation starting from the collapse of a molecular cloud have shown discs often accrete material through chaotic infall events \citep{2018Bate,2025Mayer,2026Cusack}. These sporadic accretion episodes are more likely to occur in the earliest stage of a disc's evolution. During this early phase, discs are also expected to be massive enough that the disc's self-gravity result in gravitational instability (GI), characterised by large-scale spiral structures \citep{2003bRice,2007Durisen,2016Kratter}.

Elias~2-27 \citep{2016Perez,2021Paneque} and AB~Aur \citep{2024Speedie,2025Speedie} are two discs that exhibit large-scale spiral structures which are thought to be a result of gravitational instability. Theoretical simulations of isolated gravitationally unstable discs show deviations from Keplerian rotation in the disc kinematics coincident with the spiral arms, often referred to as the \emph{GI wiggle} \citep{2020Hall,2021Longarini,2022Terry}. While the observations of the disc kinematics in Elias~2-27 and AB~Aur show global deviations from Keplerian rotation, there are additional asymmetric features that are not predicted from isolated gravitationally unstable discs. Additionally, \cite{2025bCalcino} have shown that infall alone can also reproduce the observed kinematics and large-scale spiral arms in AB~Aur. However, it does not exclude gravitational instability and infall occurring simultaneously.

Hence, we aim to investigate the impact of misaligned infall on the evolution of a gravitationally unstable disc. In particular, we study the warp induced by infall in a gravitationally unstable disc, and the impact of continuous accretion on gravitational instability. We also generate synthetic observations to explore whether the combined influence of infall and gravitational instability can produce asymmetric features as seen in observations. The 3D hydrodynamics simulations are described in \S\ref{sec:model}. We present our results and synthetic observations in \S\ref{sec:results}. We discuss the observational implications and limitations in \S\ref{sec:Discussion}. We conclude our work in \S\ref{sec:conclusion}.

\section{Methods}
\label{sec:model}

We use \textsc{Phantom} \citep{2018Price}, a smoothed particle hydrodynamics (SPH) code, coupled with \textsc{mcfost} \citep{2006Pinte,2009Pinte}, a Monte-Carlo radiative transfer code. 

\subsection{Disc Setup}

The disc is modelled using $10^6$ SPH particles between ${R_\mathrm{in} = 5 \mathrm{\ au}}$ and ${R_\mathrm{out} = 125 \mathrm{\ au}}$ in a $0.15M_\odot$ disc around a $1M_\odot$ star. The star is modelled as a sink particle \citep{1995Bate} with an accretion radius of ${5\mathrm{\ au}}$.  We set up an initial surface density profile $\Sigma = \Sigma_{0}  \left( {R}/{R_{\mathrm{in}}} \right)^{-1} f_{s}$, where ${\Sigma_{0} = 5.30 \times 10^2\,\mathrm{g\ cm}^{-2}}$,  and ${f_{s} = 1-\sqrt{R_\mathrm{in}/R}}$ smooths the surface density at the inner boundary of the disc.  The disc temperatures are determined by \textsc{mcfost}, but we set the initial vertical extent of the disc according to $H/R = 0.05 \left ( {R}/{R_{\rm in}} \right)^{0.25}$. We assume an adiabatic equation of state with an adiabatic index $\gamma=5/3$, and include self-gravity. The temperatures are calculated using \textsc{mcfost} and is described in \S\ref{sec:RT}. 

The \cite{2010Cullen} switch detects any shocks that form and generates the correct shock dissipation depending on the proximity to the shock. Near the shock, the linear shock viscosity parameter, $\alpha_{\text{AV}}$ increases up to a maximum value of 1. Further away, $\alpha_{\text{AV}}$ decreases to a minimum value of 0.1. We set the quadratic artificial viscosity parameter $\beta_{\text{AV}}=2$  \cite[][]{2015Nealon,2018Price}. The disc is resolved with an initial approximate mean smoothing length over disc scale height ${<}h{>} / H = 0.2$.

\subsection{Initialising the Infall}

\begin{figure}
    \centering
    \includegraphics[width=\linewidth]{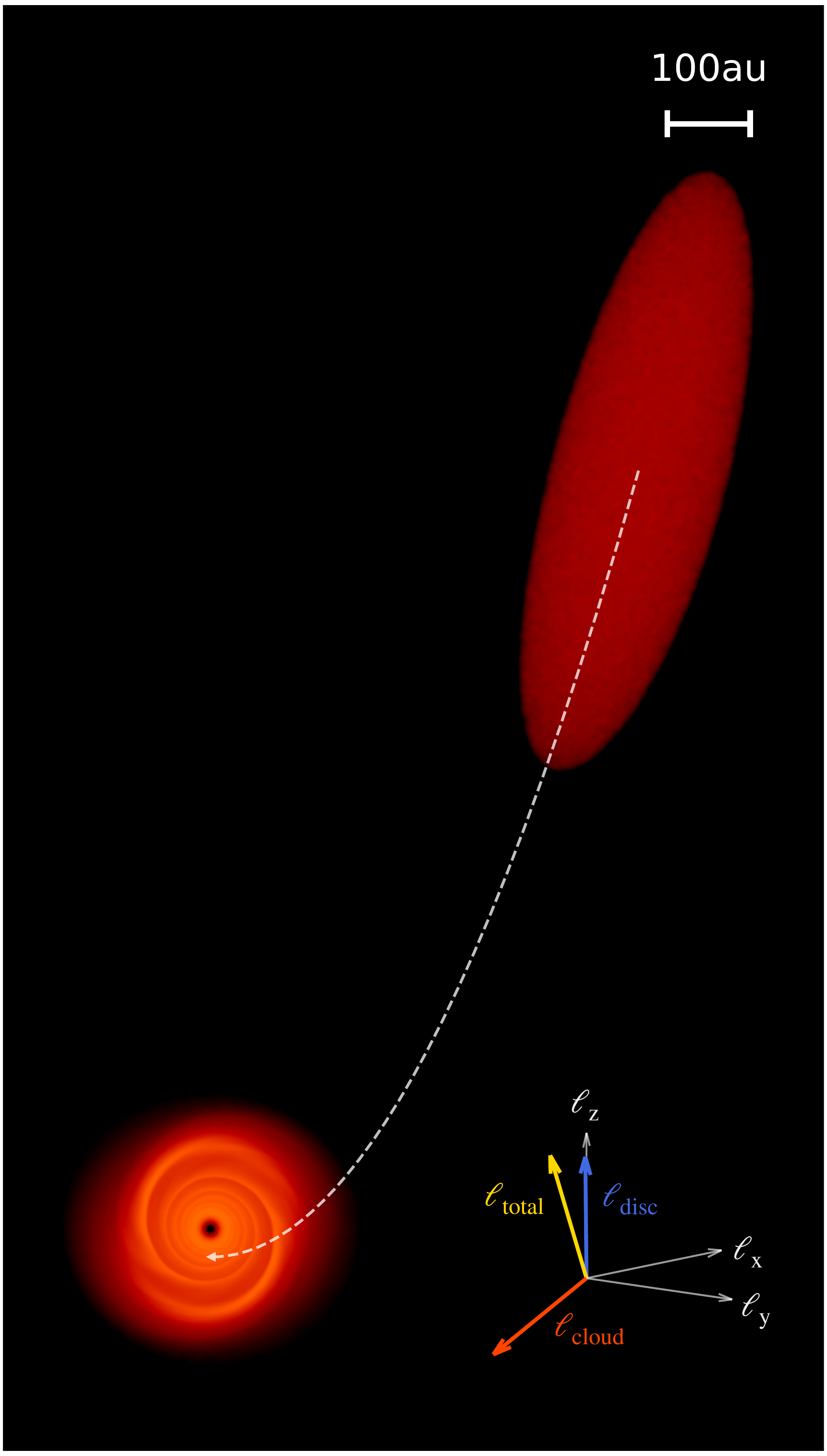}
    \caption{The simulation setup after the infalling material has been initialised. The disc has already been evolved for 4 outer orbits until gravitational instabilities developed. The infall is initialised as an elliptical cloud of gas that is one-third of the disc mass. The disc and cloud are initially misaligned by $135^\circ$ as shown by the unit angular momentum vectors in the bottom right.}
    \label{fig:init}
\end{figure}

\begin{figure*}
    \centering
    \includegraphics[width=\linewidth]{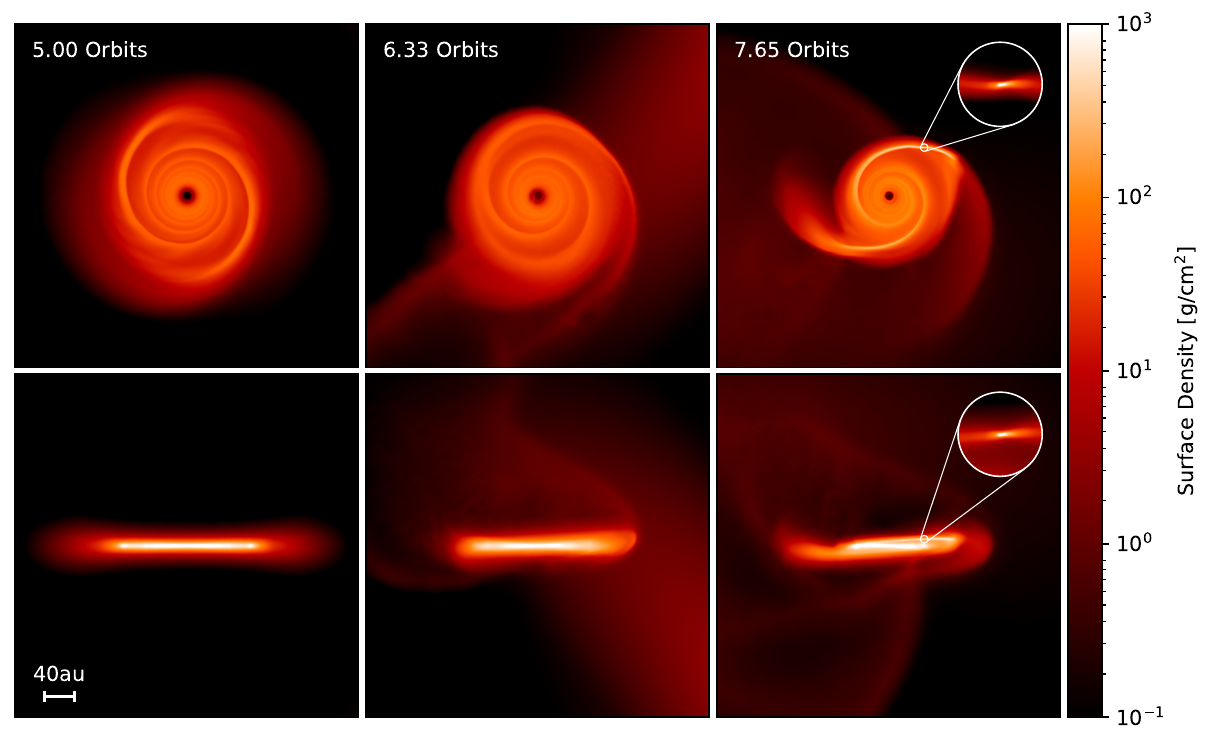}
    \caption{The evolution of the surface density of a $0.15M_\odot$ disc as it interacts with a $0.05M_\odot$ cloud of infalling gas. The top and bottom panels show the face-on and edge-on views, respectively. From left to right, the panels show the disc at 5 orbits before infall has impacted the disc, at 6.33 orbits as the disc begins to interact with the infalling gas, and at 7.65 orbits after a period of continuous infall onto the disc. Before infall, the spiral arms driven by gravitational instability are symmetric, and the disc is coplanar. As infall initially impacts the disc, the spiral structures weaken and become more asymmetric. The misaligned infall also begins to warp the disc. After a period of continuous infall, the spiral arms return to being symmetric as gravitational instability once again becomes the dominant process as the disc mass gradually increases. The warp also becomes stronger, evident in the edge-on view where the spiral arms are misaligned relative to the inner disc. Eventually, one of the spiral arms becomes gravitationally unstable and fragments to form a planet, as shown in the zoomed-in insets.}
    \label{fig:densEvol}
\end{figure*}

We first evolved the disc for 4 outer orbits until gravitational instabilities developed. Then, we initialised the infalling material as an ellipsoidal cloud of gas with an initial semi-major axis of 500 au and a semi-minor axis of 100 au. The mass of the cloud is $M_\mathrm{cloud} = 0.05M_\odot$, a third of the disc mass. The centre of the cloud is initialised at $R_\mathrm{init} = 1400$ au on a free-falling parabolic orbit as in \citet{2025Calcino}, which is similar to fly-by simulations by \citet[][see their Appendix A]{2019Cuello}. To maximise disc warping, the cloud is initially misaligned by $135^\circ$ relative to the disc \citep{2016Xiang}. The distance of closest approach is set to 50 au. The simulation is then evolved for another 3.65 orbits.

Figure \ref{fig:init} shows the simulation setup after the infalling cloud has been introduced. The dashed gray lines shows the approximate trajectory of the cloud neglecting any hydrodynamic interactions with the disc. The arrows show the initial unit angular momentum vectors, $\vb*{\ell}$ of the total system, disc \& cloud with $\vb*{\ell}_\text{disc} = [0, 0, 1]$ and $\vb*{\ell}_\text{cloud} = [0, \sfrac{-1}{\sqrt{2}}, \sfrac{-1}{\sqrt{2}}]$. Hence, the cloud approaches the disc from the negative $\vb*{z}$ direction on a retrograde, inclined orbit, with a mutual inclination of
\begin{equation}
    i = \cos^{-1}\left({\vb*{\ell}_\text{disc} \cdot \vb*{\ell}_\text{cloud}} \right)= 135^\circ.
\end{equation}

\subsection{Live Radiative Transfer}
\label{sec:RT}

The temperatures in our simulation were calculated on-the-fly using \textsc{mcfost}. We included $P\mathrm{d}V$ work and shock heating as source terms in the radiative transfer to model heating by the spiral arms and from the interaction between the disc and infalling material.

The stellar luminosity was set using the sink particle mass and a 1\,Myr isochrone from \citet{2000Siess}.  This corresponded to a stellar luminosity of $L_\star=2.06L_\odot$, temperature of $T_\star = 4280$K, and a stellar radius of $R_\star = 2.62R_\odot$. The accretion luminosity of the star is not included. As these are gas-only simulations, we assumed that the dust is perfectly coupled to the gas with a constant dust-to-gas mass ratio of 0.01. The dust grains are distributed between $0.03-1000\,\mu\mathrm{m}$ using 100 grain sizes and a power-law exponent of $\mathrm{d}n(s) \propto s^{-3.5}\mathrm{d}s$. We assumed spherical, homogeneous, astrosilicate dust grains \citep{Weingartner2001}.

Each \textsc{mcfost} Voronoi cell corresponds to an SPH particle. As in \citet{2024bRowther,2025Rowther}, we used $5 \times 10^9$ photons to illuminate the disc. Due to the high disc mass, a large ratio of photons to SPH particles is necessary to ensure photons can reach even the optically thickest regions. We have assumed radiative equilibrium, i.e., absorbed energy is instantly re-emitted. The temperatures were updated every ${1/(100\sqrt{2}) \approx 0.007071}$ of an orbit at $R_\mathrm{out}$, or every 9.88 years. We assume the gas and dust are in thermal equilibrium with $T_{\rm gas}=T_{\rm dust}$. The high frequency ensures that the particles do not undergo significant dynamical evolution between temperature updates. Thus, the particles always have the correct temperatures based on their current location.

\subsection{Synthetic Observations}

Synthetic observations were created by using \textsc{mcfost} to post-process the simulations. We assume the same stellar parameters as above, and assume a distance of 140 pc. All images were convolved with a beam of $0.05'' \times 0.05''$ (or 7 au $\times$ 7 au at 140 pc). The disc position angle is set to $-90^\circ$. The scattered light emission at $1.6\, \mu\mathrm{m}$ was generated with a disc inclination of $90^{\circ}$. The continuum emission at $1.3\, \mathrm{mm}$ was created using a disc inclination of $0$ and $90^\circ$. The molecular line emission was obtained for C$^{18}$O for the ${J=2\to1}$ transition using a disc inclination of $48^\circ$, assuming local thermodynamic equilibrium. The abundance for C$^{18}$O relative to H$_2$ was assumed to be $7 \times 10^{-7}$. We account for CO freeze-out at $T < 20\,\mathrm{K}$, and photo-dissociation and photo-desorption in regions of high UV radiation \citep[see Appendix B of][]{2018Pinte}.

\section{Results}
\label{sec:results}

\subsection{Forming a Planet in a Warped Protoplanetary Disc}

Figure \ref{fig:densEvol} shows the surface density evolution of a $0.15M_\odot$ disc as it interacts with $0.05M_\odot$ of infalling gas inclined by $135^\circ$ relative to the disc. The top and bottom panels show the face-on and edge-on views, respectively. All times are in units of orbital periods at ${R_\text{out} = 125\mathrm{\ au}}$.

The left panels show the disc at 5 orbits, before the infalling gas has perturbed the disc. In this phase, the edge-on view shows a coplanar disc where the angular momentum vector of the disc is constant with radius. The face-on view shows that the spiral structure is symmetric. The evolution of the disc is identical to that of a gravitationally unstable disc isolated from the environment. 

The middle panels show the disc at 6.33 orbits when the infalling gas begins to interact with the outer disc. As shown in Figure \ref{fig:init}, the angular momentum vectors of the disc and cloud are misaligned. As the infalling material accretes onto the disc, its misaligned angular momentum exerts a torque on the disc that alters the magnitude and direction of the disc's angular momentum vector. However, this interaction does not occur uniformly across all radii. Hence, the outer disc will experience stronger torques than the inner disc, resulting in the angular momentum vector of the disc varying with radius. By definition, the disc is warped as seen in the edge-on view. The infalling gas also induces a spiral arm which interacts with the spiral structures due to gravitational instability. The combination of these mechanisms causes the spiral arms to become more asymmetric. 

The right panels show the disc at 7.65 orbits. The infalling cloud has continuously deposited mass in the disc resulting in gravitational instability once again becoming the dominant process. The face-on view reveals that the spiral structures are stronger and have returned to being more symmetric. The edge-on view shows that the ongoing infall has also strengthened the warp, where it's now clear that the spiral arms are misaligned relative to the inner disc. Thus, the disc is simultaneously gravitationally unstable and warped. In this phase, one of the spiral arms has become gravitationally unstable and fragmented to form a planet, as shown in the zoomed-in insets. Thus the combination of misaligned infall and gravitational instability can provide a mechanism for planet formation in warped protoplanetary discs.

\subsection{Quantifying the Warp}
\label{sec:warp}

The radial variation of the disc's unit angular momentum vector, $\vb*{\ell}$ describes the degree of the warp in the disc. It can be characterised by the tilt, twist, and warp amplitude.  Therefore, we must first isolate the disc from the infalling material. A particle is considered to be part of the disc if its azimuthal velocity is near Keplerian,
\begin{equation}
    \label{eq:disc}
    \frac{\left|v_{\phi} - v_\mathrm{kep}\right|}{v_\mathrm{kep}} \leq 0.225,
\end{equation}
where $v_\mathrm{kep} = \sqrt{GM_\mathrm{enc}/R}$. The enclosed mass, $M_\mathrm{enc}$ is defined as the sum of the stellar mass and the disc mass within radius $R$, and $G$ is the gravitational constant. We then rotate the disc so that $\vb*{\ell}_z$ is parallel to the total angular momentum vector. The tilt is defined as
\begin{equation}
    \beta(R) = \arccos \left( \vb*{\ell}_z(R)\right)
\end{equation}
which measures the angle between the local unit angular momentum vector and the total angular momentum vector. The twist is defined as
\begin{equation}
    \gamma(R) = \arctan \left(\frac{\vb*{\ell}_y(R)}{\vb*{\ell}_x(R)}\right)
\end{equation}
where a change in twist over time indicates the disc is precessing. The warp amplitude is defined as
\begin{equation}
    \psi = R\left|\frac{\partial \vb*{\ell}}{\partial R}\right|
\end{equation}
which describes the steepness of the warp.

\begin{figure}
    \centering
    \includegraphics[width=\linewidth]{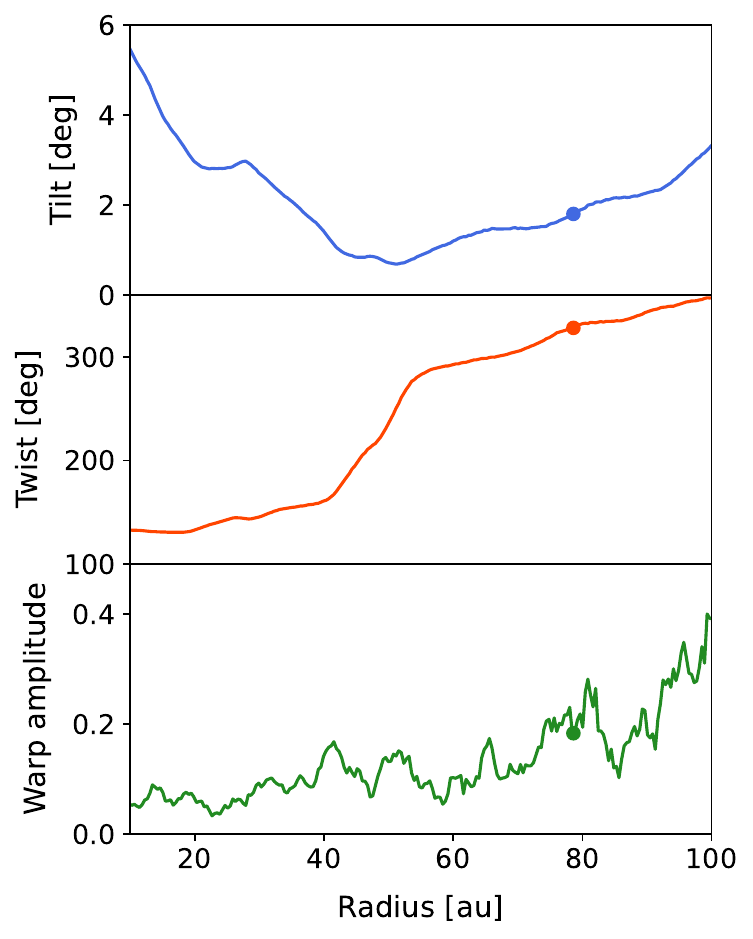}
    \caption{Radial profiles of the disc tilt (top panel), twist (middle panel), and warp amplitude (bottom panel) at 7.65 orbits. The solid circle marks the location of the fragment. The disc is strongly twisted with minor misalignments as seen by the low tilt values, but a radially dependent twist which is consistent with differential precession. The smooth increase in the warp amplitude shows that the disc remains continuous, with the strength of the warp increasing towards the outer regions of the disc.}
    \label{fig:warp}
\end{figure}

Figure \ref{fig:warp} shows the radial profiles of the tilt (top panel), twist (middle panel), and warp amplitude (bottom panel) of the disc at 7.65 orbits when the disc has fragmented (see right panels of Fig. \ref{fig:densEvol}). The solid circle marks the radial location of the fragment.

The disc is weakly misaligned with minor radial variation and tilt angles of ${\sim} 1-4^\circ$. However, there is a strong radial dependence on the twist. Beyond ${\sim} 50$ au, the twist angle is roughly $300^\circ$, whereas it decreases to roughly $140^\circ$ inside 50 au. This radial variation is consistent with differential precession. Hence, the spiral arms appear more misaligned than the tilt suggests. The apparent misalignment depends on the projection, varying with viewing angle as a consequence of the twisted nature of the disc. The warp amplitude smoothly increases from ${\sim} 0.075$ in the inner disc to ${\sim} 0.3$ in the outer disc, indicating the warp becomes stronger towards the outer regions of the disc. Therefore, the infalling material has caused the disc to become a strongly twisted, but globally continuous disc with minor misalignments.

\begin{figure}
    \centering
    \includegraphics[width=\linewidth]{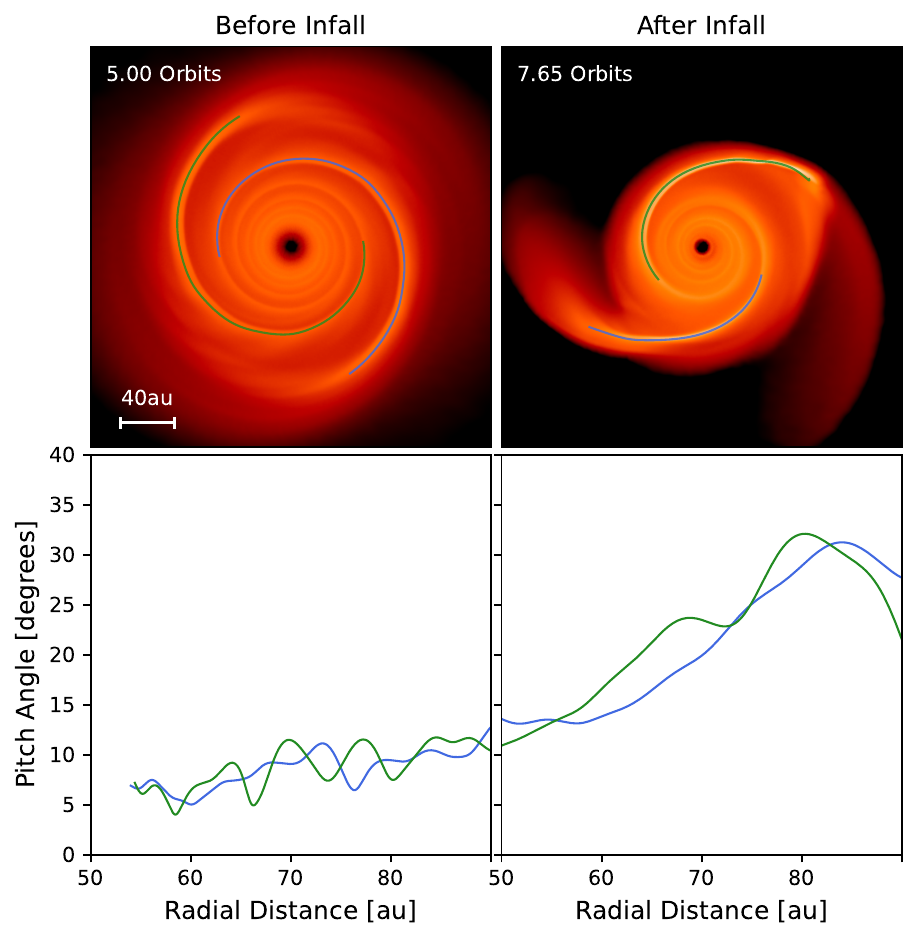}
    \caption{The top panels show the surface density of the disc at 5 (left panel) and 7.65 orbits (right panel), before and after the disc has been perturbed by infall, respectively. The extracted location of each spiral is plotted on top of the image. The bottom panels show the pitch angles of each spiral arm calculated using Eq. \ref{eq:pitchAngle}. The pitch angles before infall are constant with radius, consistent with past studies of isolated gravitationally unstable discs. However, after infall the pitch angles increase with radius.}
    \label{fig:pitchAngle}
\end{figure}

\subsection{Impact of Infall on the Spiral's Pitch Angle \& Pattern Speed}

The pitch angle, $\alpha$ at radius $R$ is the angle between the tangent of the spiral arm and the circle at $R$. It is defined as \citep{2008Binney},
\begin{equation}
    \label{eq:pitchAngle}
    \cot \alpha = \left|R \frac{\partial \phi}{\partial R}\right|.
\end{equation}
To obtain the pitch angle using Eq. \ref{eq:pitchAngle}, the locations of the spiral arms are required. We first rotate the disc such that the total angular momentum vector is aligned with the $\hat{\vb*{z}}$ direction (as in \S\ref{sec:warp}).  The two-dimensional surface density is subtracted from an axisymmetric model to obtain deviations from symmetry. We choose only the locations with positive values as they trace the rough locations of the spiral arms, which we then convert to a binary image. We then use \texttt{skeltonize} \citep{2014vanderWalt} to reduce the binary image to a 1 pixel wide representation to extract a more precise location of the spiral arms.

Figure \ref{fig:pitchAngle} compares how the spiral arms due to gravitational instability are impacted by infall. The top panels show the surface density of the disc at 5 and 7.65 orbits, respectively. The extracted locations of the two spiral arms are plotted on top of the image. The bottom panels show the pitch angles of each spiral arm calculated using Eq. \ref{eq:pitchAngle}. At 5 orbits, the disc has not yet been perturbed by the infalling gas. Hence, the spiral arms are tightly wound and symmetric with roughly constant pitch angles, consistent with expectations from isolated gravitationally unstable discs \citep{2018bForgan}. At 7.65 orbits, after the disc has been impacted by infalling gas, the disc is still gravitationally unstable with symmetric spiral arms, but the pitch angles now increase with radius. An increasing pitch angle with radius is also seen in simulations of discs after a stellar flyby \citep{2023Smallwood}.

\begin{figure}
    \centering
    \vspace{0.295cm}
    \includegraphics[width=\linewidth]{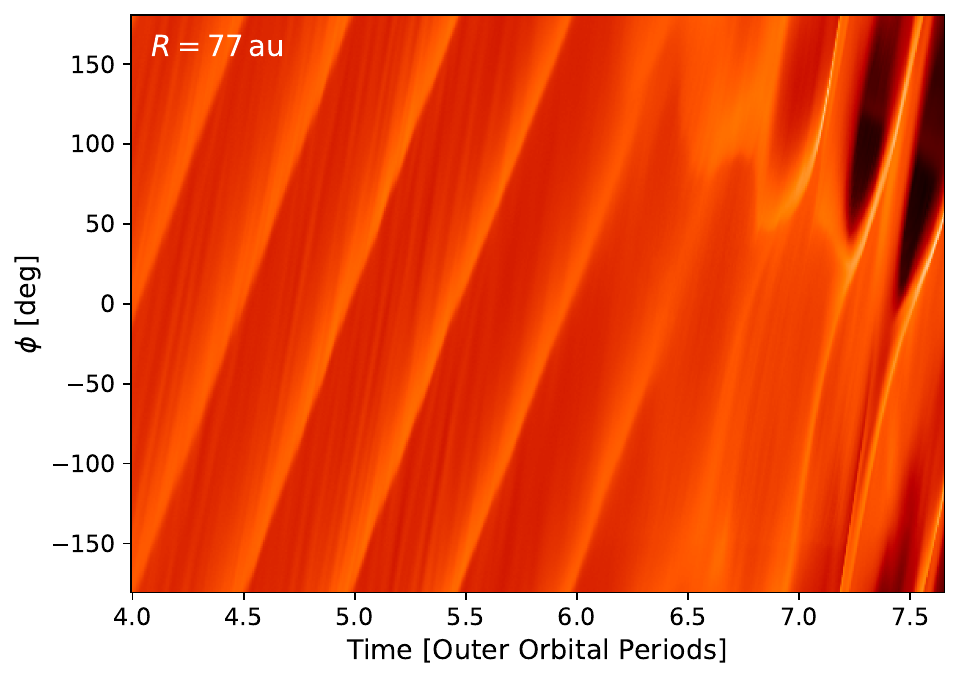}
    \caption{The evolution of the spiral arms at a fixed radius of $R=77$ au as a function of azimuthal angle $phi$, and time. Before infall (at $t < 6$ orbits), the spiral arms are straight diagonal lines, indicating a constant pattern speed at a specific radius, consistent with expectations of isolated gravitationally unstable discs. After infall, the pattern speed increases as evident by the steeper gradient. Additionally, there is a slight bend which moves towards smaller $\phi$ over time as the effects of infall extends to the entire disc.
    }
    \label{fig:spiralEvol}
\end{figure}

Figure \ref{fig:spiralEvol} shows the evolution of the surface density at a fixed radius of $R=77$ au as a function of azimuthal angle $\phi$, and time. The high density regions trace the spiral arms. Before infall ($t < 6$ orbits), the spiral arms appear as straight diagonal lines indicating a constant pattern speed at a specific radius. This is consistent with expectations from an isolated gravitationally unstable disc where the pattern speed is close to corotation \citep{2009Cossins}.

This behaviour changes at ${\sim}6.33$ orbits when infall first impacts the disc. Initially, the pattern speed of the spirals is only affected at $\phi \gtrsim 70$, coincident with where the infalling gas is initially deposited onto the disc (see upper middle panel in Fig. \ref{fig:densEvol}). The spirals retain their original pattern speed elsewhere. Over time, the gradient of the spiral arms steepens throughout the disc, indicating an increase in the pattern speed. Additionally, the gradient is no longer constant with azimuth. There is a slight bend which moves towards smaller $\phi$ over time as the impact of infall extends to the entire disc.

Thus, the combination of infall and gravitational instability results in the evolution of the spiral arms that differs from both that of an isolated gravitationally unstable disc, and that of the stationary spiral arms driven purely by infall in \citet{2025Calcino}.

\subsection{The Impact of Infall on Gravitational Instability}

\begin{figure}
    \centering
    \includegraphics[width=\linewidth]{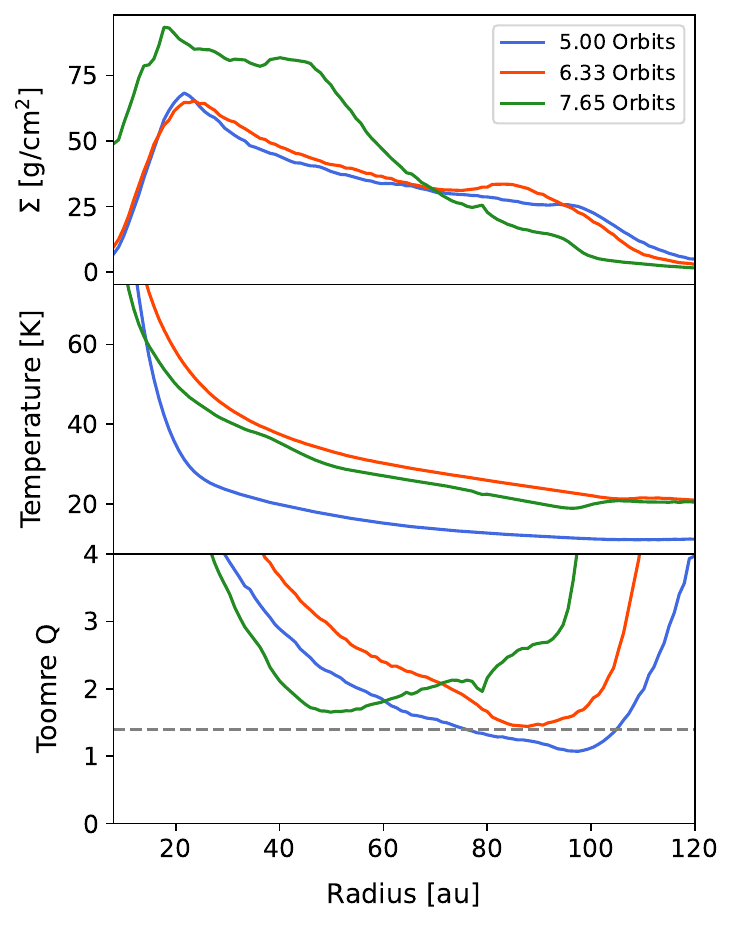}
    \caption{Azimuthally averaged profiles of the surface density (top panel), temperature (middle panel), and Toomre $Q$ (bottom panel) at 5, 6.33, and 7.65 orbits. The infalling gas causes mass to be redistributed from the outer disc towards the inner disc. Additionally, the disc temperatures increase due to envelope backwarming. Although Toomre $Q$ indicates that the disc should become gravitationally stable as a result, this is a limitation of azimuthally averaging.}
    \label{fig:Q}   
\end{figure}

\begin{figure}
    \centering
    \includegraphics[width=\linewidth]{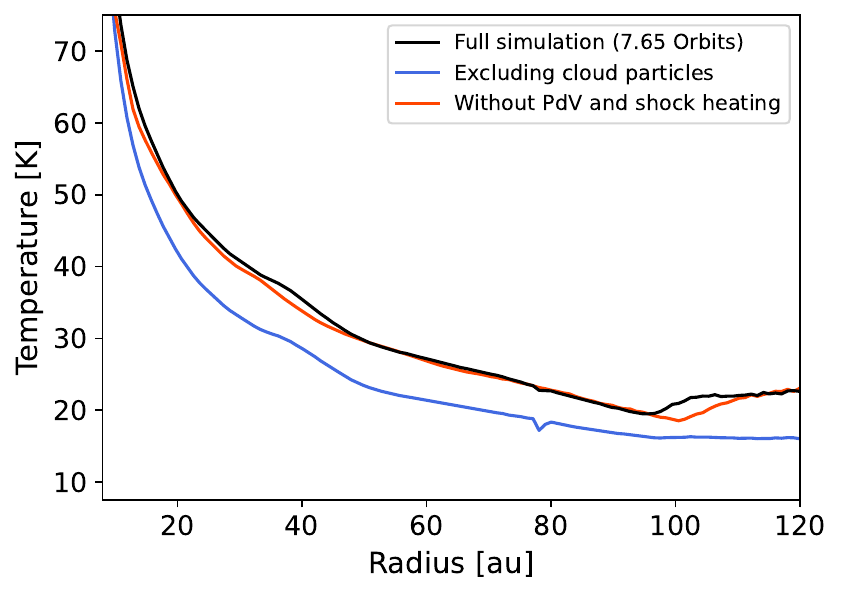}
    \caption{All three lines show the azimuthally averaged temperature at 7.65 orbits. The black line uses all particles in the simulations and includes $P\mathrm{d}V$ work and shock heating. The blue line shows the temperature recalculated by \textsc{mcfost} after excluding the cloud particles. The orange line shows the temperatures obtained without including $P\mathrm{d}V$ work and shock heating. There are minor differences in the temperatures without $P\mathrm{d}V$ work and shock heating. However, the disc is cooler when the cloud particles are removed. Thus, radiative interactions with the surrounding infalling gas is the dominant reason for the disc becoming warmer.}
    \label{fig:temperature}
\end{figure}

\begin{figure}
    \centering
    \vspace{-0.025cm}
    \includegraphics[width=\linewidth]{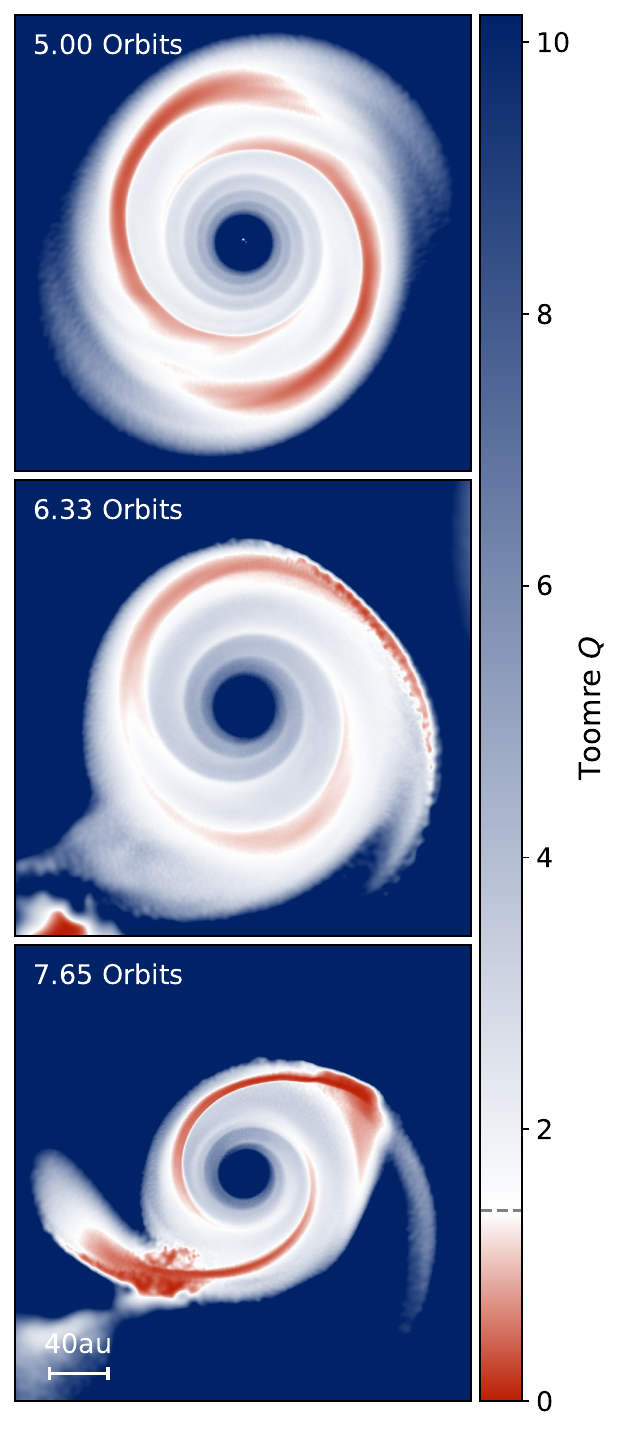}
    \caption{Two-dimensional plots of Toomre $Q$ at 5, 6.33, and 7.65 orbits. The grey dashed line marks ${Q = 1.4}$. At 5 orbits, pre-infall, the low $Q$ regions in red correspond to the spiral arms as expected. At 6.33 orbits, most of the disc has ${Q > 1.4}$, consistent with the azimuthally averaged profile in Fig \ref{fig:Q}. However, at the location where infall first impacts the disc (top right quadrant), there is a region of low $Q$, highlighting the localised nature of infall. At 7.33 orbits, the spiral arms once again correspond to low $Q$. However, they are messier due to the continued effect of infall. Thus, gravitational instabilities are still visible due to local conditions even when the azimuthally averaged criterion, ${Q > 1.4}$ is not satisfied.}
    \label{fig:2dQ}
\end{figure}

Figure \ref{fig:Q} plots the azimuthally averaged surface density (top panel), temperature (middle panel), and \citet{1964Toomre} stability parameter $Q$ (bottom panel) at 5, 6.33, and 7.65 orbits. A disc is expected to become gravitationally unstable and form spiral arms when \citep{2007Durisen}
\begin{equation}
    \label{eq:Toomre}
    Q = \frac{c_{\text{s}} \Omega}{\pi G \Sigma} \lesssim 1.4,
\end{equation}
where $c_{\text{s}}$ is the sound speed, $\Omega$ is the orbital frequency, and $\Sigma$ is the surface density. The grey dashed line marks ${Q = 1.4}$. The disc is coldest at 5 orbits before it has been perturbed by infalling gas. In the outer regions of the disc, ${Q < 1.4}$, consistent with a gravitationally unstable disc.

\begin{figure}
    \centering
    \includegraphics[width=\linewidth]{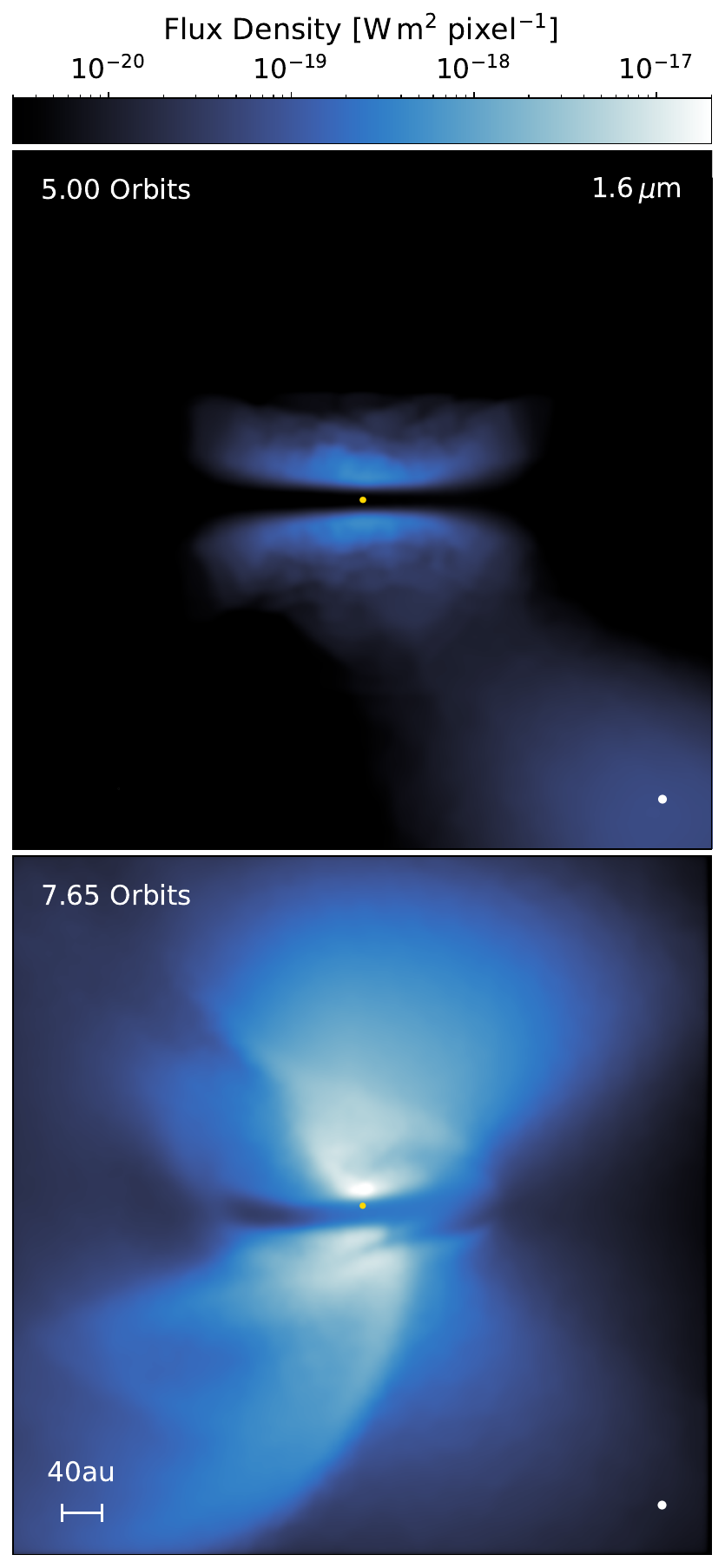}
    \caption{Synthetic images of the disc at $1.6\,\mu$m when viewed edge-on at 5 and 7.65 orbits. Since the disc is coplanar before infall (top panel), the flux from the upper disc surfaces is symmetric. After infall (bottom panel), there is a lateral asymmetry in the flux due to the disc being warped.}
    \label{fig:micron}
\end{figure}

\begin{figure}
    \centering
    \includegraphics[width=\linewidth]{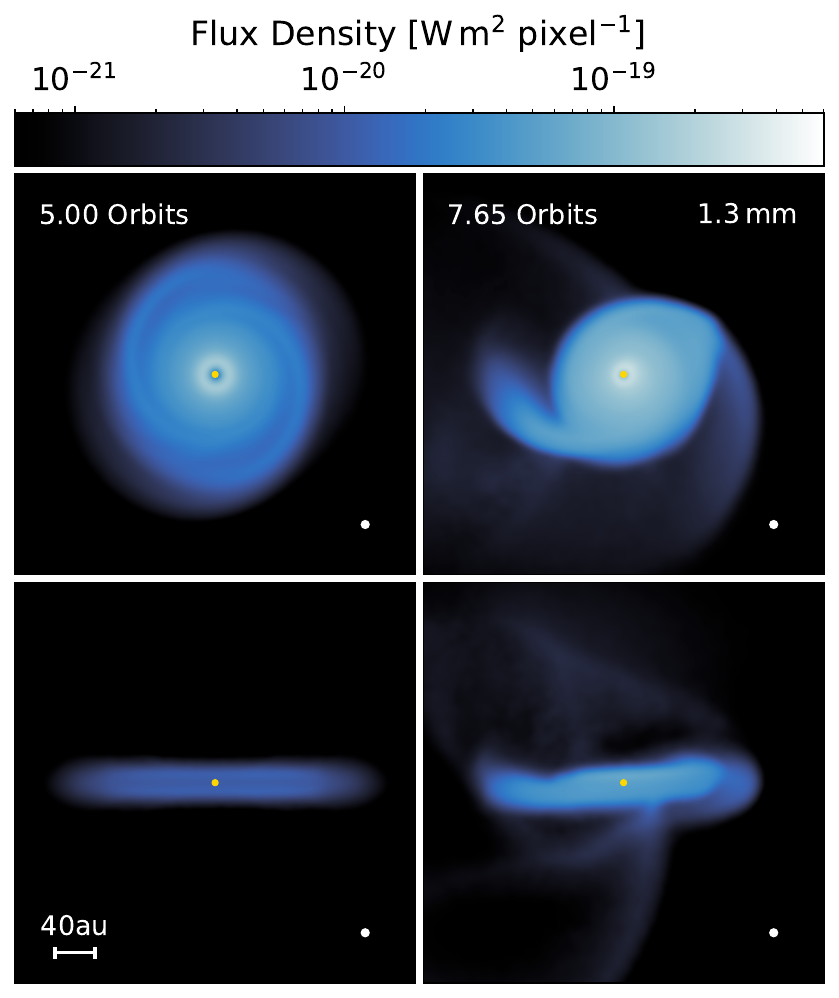}
    \caption{Synthetic images of the disc at $1.3 \mathrm{\, mm}$ at 5 and 7.65 orbits when viewed face-on and edge-on are shown in the top and bottom panels, respectively. Infalling gas increases the disc mass and decreases the radial extent of the disc. Thus, the contrast of the spiral arms is lower due to the increased optical depth of the disc. The warp induced by infall is also evident in the edge-on view where the misaligned spiral arms result in vertical asymmetries. The infalling gas itself is also visible in both the face-on and edge-on view.}
    \label{fig:mm}
\end{figure}

At 6.33 orbits, the azimuthally averaged disc temperatures have increased by roughly 20K. There is also a slight increase in the surface density, but not enough to offset the increase in temperature, as seen by an increase in $Q$ pushing the disc towards stability. This is also consistent with the top middle panel in Fig. \ref{fig:densEvol}, where the spiral arms are weaker compared to at 5 orbits. At 7.65 orbits, after the initial impact of infall, the disc temperatures has a slight decrease, but remains roughly 15K higher globally compared to the pre-infall phase. The reason for the increased temperatures can be seen in Figure \ref{fig:temperature}, which compares the relative importance of the surrounding cloud and $P\mathrm{d}V$ work \& shock heating on the disc temperature. The disc is cooler when the temperatures are recalculated after excluding the cloud particles. Neglecting $P\mathrm{d}V$ work and shock heating has a negligible impact on the disc temperature. Thus, reprocessing of stellar irradiation by the surrounding infalling gas (known as envelope back-warming) is responsible for the disc becoming warmer \citep{1994Butner,1997DAlessio,2023Han}.

At 7.65 orbits, the surface density in the outer disc has slightly decreased, as mass is redistributed inwards. The azimuthally averaged $Q$ indicates that the disc should be gravitationally stable given ${Q > 1.4}$ throughout out the disc. However, the disc is clearly gravitationally unstable as from Fig. \ref{fig:densEvol}, the spiral arms are stronger, and one of the arms has fragmented to form a planet. The reason for this discrepancy is because infalling gas is asymmetric, leading to localised changes in the surface density. Thus, gravitational instabilities are determined by local conditions.

\begin{figure*}
    \centering
    \includegraphics[width=\linewidth]{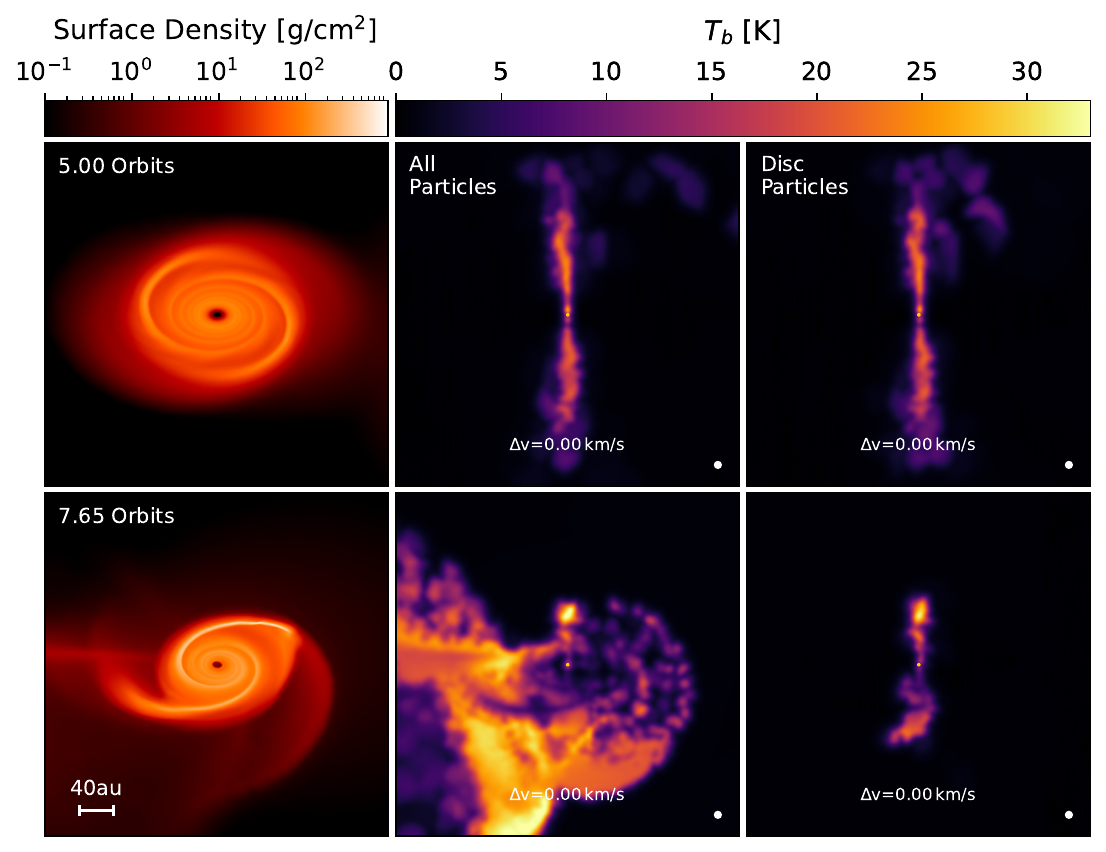}
    \caption{Comparison of the the iso-velocity curves of C$^{18}$O synthetic observations for the ${J=2\to1}$ transition at $\Delta v = 0$. The left panels show the surface density of the disc, viewed at an inclination of $48^{\circ}$ for consistency with the channel maps in the middle and right panels. The middle panels use all particles in the simulation to obtain the channel maps. The right panels only use the disc particles selected using Eq. \ref{eq:disc}. The top panels show the disc in the pre-infall phase at 5 orbits. The bottom panels show the disc at 7.65 orbits when it is both gravitationally unstable and warped by ongoing infall. At 5 orbits, there are global deviations from a smooth Keplerian rotation, consistent with gravitational instability. At 7.65 orbits, the channel maps are messier due to infall. A better understanding is obtained by isolating the disc particles. It is clear that most of the features seen in the lower-middle panel are due to infall. In contrast to the pre-infall phase, the emission from the disc is bendy and asymmetric, which explains the apparent lack of emission in the bottom half of the disc in the lower-middle panel.}
    \label{fig:13C16O}
\end{figure*}

To illustrate this, Figure \ref{fig:2dQ} shows the two-dimensional Toomre $Q$ parameter at 5, 6.33, and 7.65 orbits. The grey dashed line marks $Q = 1.4$. At 5 orbits, before infall, the spiral arms correspond to low $Q$ values (represented by red), consistent with the azimuthally averaged profile. At 6.33 orbits, the disc is more stable with ${Q > 1.4}$ through most of the disc, consistent with the warmer temperatures seen in Fig. \ref{fig:Q}. However, there is a low $Q$ region in the top right quadrant, coincident with where the infalling gas first impacts the disc. The mass in the spiral arms gradually increases as infalling gas continues to accrete onto the disc. At 7.65 orbits, gravitational instabilities are stronger as seen by the low $Q$ values that correspond to the spiral arms. However, unlike the pre-infall phase, $Q$ is messier due to the effects of ongoing infall. Thus, highlighting that infalling material results in the disc becoming locally gravitationally unstable, despite most of the disc having ${Q > 1.4}$. This continuous increase in the mass in the spiral arms due to infall causes the disc to fragment despite the disc becoming hotter.

Thus, infall has two opposing influences on gravitational instability. The continuous deposition of mass onto the disc makes the disc more gravitationally unstable. However,  there is also a stabilising influence due to the surrounding environment re-processing stellar radiation back on to the disc, increasing the disc temperature.

\begin{figure}
    \centering
    \includegraphics[width=\linewidth]{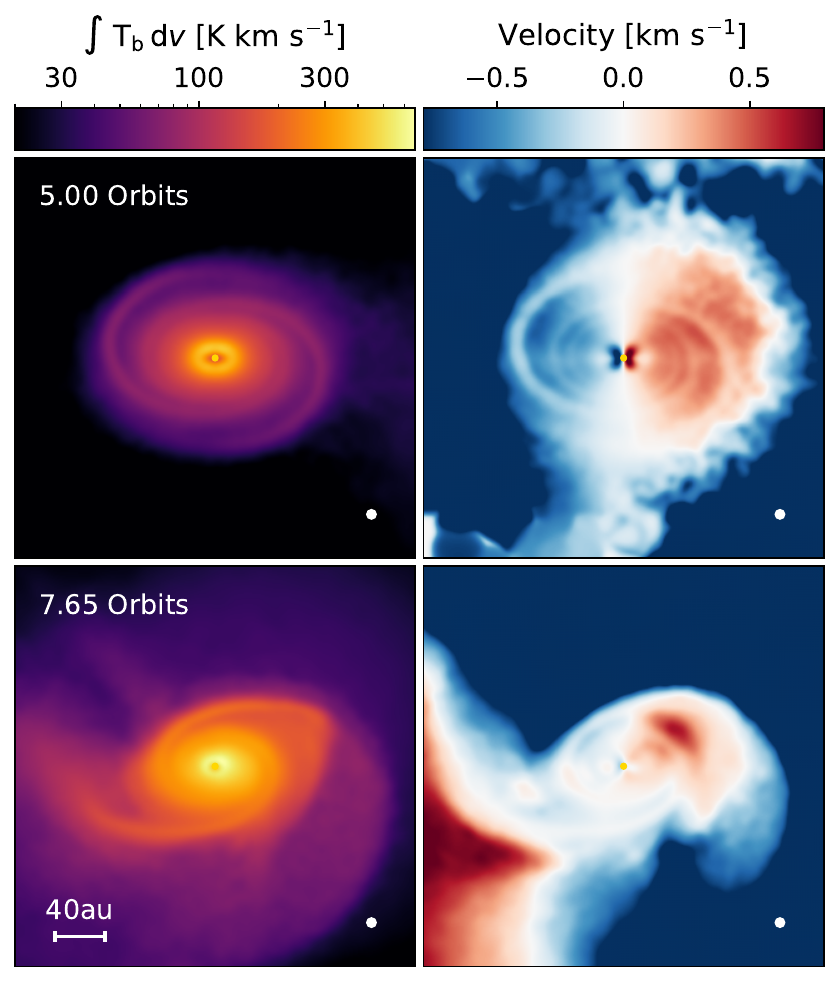}
    \caption{Comparison of the molecular line emission C$^{18}$O for the ${J=2\to1}$ transition at 5 and 7.65 orbits. The velocity-integrated brightness (moment 0) and intensity-weighted velocity (moment 1) maps are shown in the left and right panels, respectively. At 5 orbits, the spiral arms driven by gravitational instability are visible in both the moment 0 and moment 1 maps. At 7.65 orbits, the bright red spot in the top right quadrant of the moment 1 map reveals where the infalling material impacts the disc. However, the spiral arms are no longer visible in the moment 1 map. The emission in the moment 0 map is messier after infall has interacted with the disc, but the spiral arms remain visible.}
    \label{fig:momentMaps}
\end{figure}

\subsection{Synthetic Observations}

\subsubsection{Asymmetries in Edge-on Scattered Light Images}

Figure \ref{fig:micron} shows synthetic images at $1.6 \mu$m of the disc at 5 and 7.65 orbits when viewed edge-on, in the top and bottom panels, respectively. At 5 orbits, the disc is coplanar since infall has not yet impacted the disc. Hence, the flux from the upper and lower disc surfaces is symmetric. The flux seen in the bottom right is the infalling gas prior to its interaction with the disc. At 7.65 orbits, the disc is warped due to misaligned infall, resulting in a lateral asymmetry in the flux, consistent with expectations of a warped disc \citep{2019Nealon,2024Villenave,2025Kimmig}.

There is also a difference in the flux brightness between the upper and lower surfaces. While this can be indicative of a warp \citep{2019Nealon}, this difference can also occur if the disc is not perfectly edge-on. 
As seen in Fig. \ref{fig:warp}, the disc is slightly tilted after infall. The synthetic image was also re-calculated using a viewing inclination of $92^\circ$ so that the outer disc was perfectly edge-on. However, the observed asymmetries were unchanged. Thus, the asymmetries can be attributed to the warped structure rather than the small tilt of the disc.

\subsubsection{1.3mm Continuum Images}

Figure \ref{fig:mm} shows synthetic images at $1.3\,\mathrm{mm}$ of the disc at 5 and 7.65 orbits. The top and bottom panels show the face-on and edge-on views, respectively. At 5 orbits, the spiral arms are clearly visible in the face-on view with high contrast between the spirals and the inter-arm regions. As the disc is coplanar, the disc is symmetric when viewed edge-on. At 7.65 orbits, the optical depth of the disc is larger due to infalling gas increasing the disc mass and decreasing the radial extent of the disc (see top panel in Figure \ref{eq:Toomre}). Hence, the contrast between spiral arms driven by gravitational instability inside 100 au and the inter-arm regions has decreased, making the spiral arms less visible. As the disc is now warped, the edge-on view reveals asymmetries due to the spiral arms being misaligned relative to the inner disc. Infalling gas is also visible in both the face-on and edge-on views. However, it should be noted that the flux at 1.3mm (i.e., ALMA Band 6) may be slightly underestimated due to the maximum dust size in \S\ref{sec:RT} being limited to 1\,mm.

\subsubsection{Channel Maps}

Figure \ref{fig:13C16O} compares the iso-velocity curves of C$^{18}$O synthetic observations for the ${J=2\to1}$ transition at $\Delta v = 0$. The left panels show the surface density of the disc, viewed at an inclination of $48^{\circ}$ for consistency with the channel maps in the middle and right panels. The top panels show the disc in the pre-infall phase at 5 orbits. The bottom panels show the disc at 7.65 orbits when it is both gravitationally unstable and warped by ongoing infall.

The middle panels show the channel map obtained using all particles in the simulation. At 5 orbits, there are clear deviations from smooth Keplerian rotation. These are visible at all radii and are caused by the spiral arms. The morphology is consistent with previous studies of isolated gravitationally unstable discs \citep{2020Hall,2021Longarini,2022Terry}. This is unsurprising since infall has not yet perturbed the disc, and thus the kinematics are identical to an isolated disc. However, at 7.65 orbits, the channel maps are dominated by the ongoing infall. It is also difficult to distinguish between the emission from the disc and the infalling gas. The reason for the lack of emission from the disc in the bottom half is also unclear.

To gain a better understanding of the channel maps, we eliminate the emission from the cloud. To do this, we isolate the disc from the cloud by separating particles based on their azimuthal velocity. Any particle that satisfies Eq. \ref{eq:disc} with roughly Keplerian azimuthal velocity is considered as part of the disc. The right panels show the channel map obtained by using only the disc particles. At 5 orbits, the emission is unchanged, confirming the above criterion does not exclude genuine disc particles. At 7.65 orbits, the impact of infall is clear. Most of the emission seen in the lower-middle panel is due to the infalling material. Removing the cloud particles reveals the emission from the disc is now asymmetrical due to infall causing localised perturbations to the disc velocity field. This is evident by the bottom `wing' appearing more bent. Excluding the cloud particles also removes the effect of backwarming. However, this only reduces the brightness temperature and does not impact the morphology of the disc emission.

The distorted and asymmetric wings are roughly comparable to the channel maps in observations such as Elias 2-27 which is thought to have infall induced gravitational instabilities and potentially also warped \citep{2021Paneque}. However, the emission from the right panels of Fig. \ref{fig:13C16O} is not directly comparable since the contribution from the infalling cloud cannot simply be removed in actual observations. Longer-term simulations are required to determine whether the asymmetries from the disc emission are still visible after the cloud no longer contaminates the emission.

\subsubsection{Moment Maps}

Figure \ref{fig:momentMaps} shows the velocity-integrated brightness temperature (moment 0) and the intensity-weighted velocity (moment 1) maps of C$^{18}$O for the ${J=2\to1}$ transition in the left and right panels, respectively. The moment maps are shown for the disc at 5 and 7.65 orbits in the top and bottom panels, respectively. 

At 5 orbits, the spiral arms driven by gravitational instability are visible in both the moment 0 and moment 1 maps. The red-blue split is to due to the rotation of the disc when viewed at an inclined angle. The blue and red regions show the gas moving towards and away from the observer, respectively. At 7.65 orbits, the moment 0 map is messier due to infalling material, but the spiral arms are still visible. However, infall affects the moment 1 map to a greater extent. The spiral arms are no longer visible, and there is a bright red spot in the top right quadrant where the infalling material impacts the disc. While the red-blue split is still visible, it is less symmetric compared to the pre-infall phase due to the warp \citep{2015Casassus,2020Bi}.

\section{Discussion}
\label{sec:Discussion}

\subsection{The General Impact of Infall on Disc Warping}

The extent to how warped the disc becomes depends on the angular momentum of the cloud relative to that of the disc. In this work, as seen from Figure \ref{fig:init}, the disc dominates the angular momentum budget. Hence, although the trajectory of the cloud is chosen to maximise warping, a retrograde orbit inclined by $135^\circ$ \citep{2016Xiang}, it is not surprising that the misaligned infall only resulted in a minor warp with small misalignments. A more massive cloud with larger angular momentum would exert stronger torques on the disc, potentially resulting in more severe warps with larger misalignments.

The simulation becomes too computationally expensive to evolve further once the disc undergoes fragmentation. This limits our ability to determine the extent to which infall can warp a disc since infall is still ongoing at 7.65 orbits, the final snapshot. Therefore, it is likely that if the simulation was evolved further, infall would continue to torque the disc, potentially resulting in stronger warps, or even a misalignment between the fragment and the disc. 

The evolution of the fragment itself will further complicate how infalling gas warps the disc. As the fragment grows to form a giant planet or brown dwarf, it will open up a gap in the disc \citep{2015Malik,2018Stamatellos,2020bRowther,2022bRowther}. Misaligned infall will continue to torque the outer disc. However, realignment between the inner and outer disc will be less efficient since the gap limits angular momentum transfer between the inner and outer disc. Thus, misaligned infall onto a disc with a wide gap could result in larger misalignments \citep{2018Nealon}.

\subsection{The General Impact of Infall on Gravitational Instability}

As infall accretes onto the disc, it has two opposing influences on gravitational instability. The surrounding infalling gas reprocesses stellar irradiation and re-emits it back towards the disc, increasing the disc temperatures. Opposing this stabilising influence is the continuous supply of infalling gas which increases the disc mass making it more gravitationally unstable.

This competition is closely related to the infall-regulated GI scenario of \citet{2025Longarini}. They showed that continuous mass injection can drive discs towards marginal stability by replenishing the disc mass rather than through the balance between radiative cooling and spiral-shock heating \citep[e.g.][]{2001Gammie, 2004Lodato, 2016Kratter}. Our simulation agrees with the mass-loading aspect of this picture, although in our scenario the mass-loading is sufficient to drive fragmentation. In addition, since we calculate the radiation field live, the same infalling material also heats the disc through envelope back-warming \citep{1994Butner,1997DAlessio,2023Han}. Thus the infalling material not only increases the surface density, but can also increase the disc temperature. However, in the general case the relative importance of the stabilising increase in temperature and the destabilising mass growth is unclear, as it depends on the disc and cloud properties. If a lower mass cloud generated a similar rise in disc temperatures, then gravitational instability would weaken as a result of reduced mass growth by infall.

Spiral arms can also be formed by the interaction between infalling material and the disc \citep{2025bCalcino,2025Calcino,2026Huhn}, an effect that occurs in both gravitationally unstable discs \citep{2010Kratter,2011Harsono,2025Longarini} and non-self-gravitating discs \citep{2015Lesur,2016Hennebelle}.  Infalling gas is non-Keplerian, so the interaction between the infalling material and the (Keplerian) disc creates a strong vertical gradient in the azimuthal velocity. This is unstable to the Kelvin-Helmholz / Rossby-wave instability \citep{2018Bae, 2022Kuznetsova}, and for strong infall the disc can also become Rayleigh unstable \citep{2015Lesur}; both instabilities act to drive spiral density waves. Here the effect of these spirals is to change the character of the gravitational instability \citep[as in][]{2011Harsono}, but if the disc mass was lower, infall-driven spirals may form instead.

If a warp is sufficiently large, it alone can weaken gravitational instabilities. However, since the warp induced by infall is small (see Fig. \ref{fig:warp}), with a similar warp amplitude to the weakest warp in \citet{2022Rowther}, the warp itself does not have an impact on the evolution of gravitational instability.

\subsection{Limitations}

Elias 2-27 \citep{2016Perez,2021Paneque} and AB~Aur \citep{2024Speedie,2025Speedie} are both thought to be gravitationally unstable due to their spiral structures, and also impacted by infall. While this is qualitatively similar to the simulation described in this work, there are numerous differences in the disc, cloud \& stellar parameters, along with simplifying assumptions that make direct comparisons difficult.

The size of the gas disc in this simulation (${\sim}100\,\mathrm{au}$) is significantly smaller than both of these discs. An order of magnitude smaller in the case of AB~Aur. The difference in size means that the optical depth in our simulations is much higher. Thus, the observability of the spirals in scattered light and continuum emission cannot be directly compared. The stellar luminosity in our simulation ($2.06\,L_\odot$) is also different compared to the stellar luminosity in Elias~2-27 ($0.91\,L_\odot$) and AB~Aur $(59\,L_\odot)$. Combined with the different disc size, the molecular line emission is also difficult to compare due to the sensitive dependence on the temperature, which depends on the stellar luminosity and disc size. 

A caveat to the simulation presented here is that these are gas-only simulations. For the radiative transfer calculations, we have assumed the dust is perfectly coupled to the gas. While this greatly reduces the computational expense of the simulation, this assumption is invalid. Observations of discs such as Elias 2-27 \citep{2016Perez,2021Paneque} and AB Aur \citep{2024Speedie,2025Speedie} reveal that the large millimeter-sized grains are more radially compact than the gas disc. 

Additionally, we make another simplifying assumption that the dust size distribution is identical in the cloud and the disc.  This assumption is also invalid, since both of these discs are thought to be gravitationally unstable due to their spiral structures. Dust growth is expected to be much more efficient in the disc compared to the cloud due to the spiral arms which can efficiently trap dust \citep{2006Rice,2012Gibbons,2022Baehr,2023bLongarini,2024Rowther,2025Rice}.

Replicating these discs would require using comparable disc, star, and cloud properties, along with simulations with both dust \& gas. Hence, in this work, we have instead made more general predictions of how infall impacts a gravitationally unstable disc by contrasting it with an isolated disc. In particular, how the disc features become more asymmetric after infall. These include lateral asymmetries in scattered-light emission due to the disc being warped, vertical asymmetries in continuum emission due to misaligned spiral arms, and a bright red spot in one quadrant in the moment 1 map where infalling gas impacts the disc. 

\subsection{Observational Implications}

Although we do not aim to model any specific disc, our simulation of ongoing infall in young discs has several observational implications.

Our synthetic scattered-light calculations show that infall generates asymmetric and vertically extended structure through multiple mechanisms. Infall continuously adds mass to the outer disc, but this material is deposited non-uniformly onto the disc, resulting in the density structure becoming asymmetric and vertically extended. The misaligned angular momentum of infalling gas also warps the disc, causing the orientation of the disc surface to vary with radius, which in turn results in lateral asymmetries in the flux. This is consistent with recent observations of protoplanetary discs which have revealed large-scale asymmetries in scattered-light emission \citep{2024Garufi,2024Villenave}.

One method to estimate the mass of the disc is through its kinematics. In a sufficiently massive disc, self-gravity alters the rotational velocity of the disc, and this can be used to infer the disc's mass \citep{2021Veronesi,2023Lodato,2024Veronesi,2025bLongarini,2026Pezzotta}. However, this method assumes the disc is isolated from the star-forming environment, so that any deviations from Keplerian rotation can be attributed to the disc's self-gravity. The presence of infall complicates the analysis of gravitational instability, particularly when that analysis depends on the disc kinematics \citep{2025bCalcino}. Methods of estimating disc masses that do not \emph{a priori} account for the infall are likely to be biased towards higher disc mass estimates, since infall also causes deviations from Keplerian rotation. Thus, infall should not be neglected in discs such as Elias~2-27 \citep{2016Perez,2021Paneque} and AB~Aur \citep{2024Speedie,2025Speedie}, where infall and gravitational instability could be acting in unison.

Our results also show that the star-forming environment impacts the thermal structure of young, gravitationally unstable discs. The surrounding material absorbs and re-emits stellar irradiation back onto the disc, warming it up \citep{1994Butner,2023Han}. Thus, the evolution of gravitational instabilities is determined by the competing effects of the destabilising influence of increasing disc mass, and the stabilising influence of the surrounding gas reprocessing stellar irradiation. These results emphasise that the interactions between young discs and the star-forming environment cannot be neglected when interpreting observations.

\section{Conclusions}
\label{sec:conclusion}

We use 3D smoothed particle hydrodynamics simulations to investigate the impact of misaligned infall on the evolution of a gravitationally unstable disc. Our main results are:
\begin{enumerate}
    \item Misaligned infall can drive planet formation in warped discs via gravitational instability.
    \item Since the angular momentum vector of the infalling gas is oriented in a different direction relative to the disc, the infalling gas exerts a torque as it accretes onto the disc. This alters the direction of the disc's angular momentum vector. However, the interaction between infall and the disc does not occur equally across all radii. The torques are strongest in the outer regions, where the infalling gas first interacts with the disc. Hence, the disc develops a radially varying angular momentum vector, and is therefore warped.
    \item 
    Absorption and re-emission of stellar radiation by the infalling gas heats the disc significantly, weakening the spiral arms driven by gravitational instabilities. The infalling gas also continuously increases the mass in the spiral arms. Over time, the spiral arms strengthen as gravitational instability once again becomes the dominant process. Eventually, one of the spiral arms fragments, forming a planet.
\end{enumerate}

We also post-process the simulations to generate synthetic observations of the disc at $1.6\,\mu$m and $1.3\,\mathrm{mm}$, as well as molecular line emission from C$^{18}$O for the ${J=2\to1}$ transition. Our main findings about how misaligned infall impacts observations of gravitationally unstable discs are:
\begin{enumerate}
    \item As a consequence of the disc being warped by infall, there is a lateral asymmetry in the scattered-light emission from the upper and lower disc surfaces when viewed edge-on at $1.6\,\mu$m.
    \item Infalling gas increases the disc mass and decreases the radial extent of the disc. Both of these effects increase the optical depth. Thus, the spiral arms have less contrast when viewed face-on at $1.3\,\mathrm{mm}$. There are vertical asymmetries when viewed edge-on due to the misaligned spiral arms.
    \item The channel map at $\Delta v = 0$ is messier and difficult to interpret after infall impacts the disc. By isolating the disc particles, it becomes clear that most of the emission is due to infalling gas. The disc emission also becomes bent and asymmetric, in contrast to the straight morphology from an isolated disc.
    \item The spiral arms remain visible in the moment 0 map. However, instead of the spiral arms, the main feature in the moment 1 map is a bright red spot where the infalling gas impacts the disc. 
\end{enumerate}

\section*{Acknowledgements}
SR would like to thank Farzana Meru, Rebecca Nealon, and Daniel Price for interesting discussions.
SR and RA acknowledge support funding from the Science \& Technology Facilities Council (STFC) through Consolidated Grant ST/W000857/1. SR also acknowledges funding from The Royal Society. JC acknowledges support by the National Natural Science Foundation of China under grant No. 12233004, 12250610189, and W2533008. This work was performed using Avon, the HPC clusters at the University of Warwick, the computing facilities provided by the Scientific Computing Research Technology Platform of the University of Warwick, the ALICE High Performance Computing Facility and the DiRAC Data Intensive service (DIaL), managed by the University of Leicester Research Computing Service on behalf of the STFC DiRAC HPC Facility (www.dirac.ac.uk). The DiRAC service at Leicester was funded by BEIS, UKRI and STFC capital funding and STFC operations grants. DiRAC is part of the UKRI Digital Research Infrastructure. 

\section*{Data Availability}

Videos and select snapshots of the simulations along with the data to recreate any of the simulations or the figures in the paper are available at 
\url{https://doi.org/10.5281/zenodo.24289395}. We utilised the following public software:\\

{\hypersetup{urlcolor=auburn}

\begin{tabular}{ll}
\hspace{-0.52cm}\textsc{Phantom}
&\url{https://github.com/danieljprice/phantom}\\
& \citep{2018Price} \\
\hspace{-0.52cm}\textsc{mcfost} &\url{https://github.com/cpinte/mcfost}\\
& \citep{2006Pinte,2009Pinte} \\
\hspace{-0.52cm}\textsc{Splash}
&\url{https://github.com/danieljprice/splash}\\ & \citep{2007Price} \\
\hspace{-0.52cm}\textsc{Sarracen} &\url{https://github.com/ttricco/sarracen}\\
& \citep{sarracen}
\end{tabular}

}


\bibliographystyle{mnras.bst}
\bibliography{GI_infall.bib} 
\vfill
\appendix

\section{Isolating the Disc}

\begin{center}
\includegraphics[width=\columnwidth]{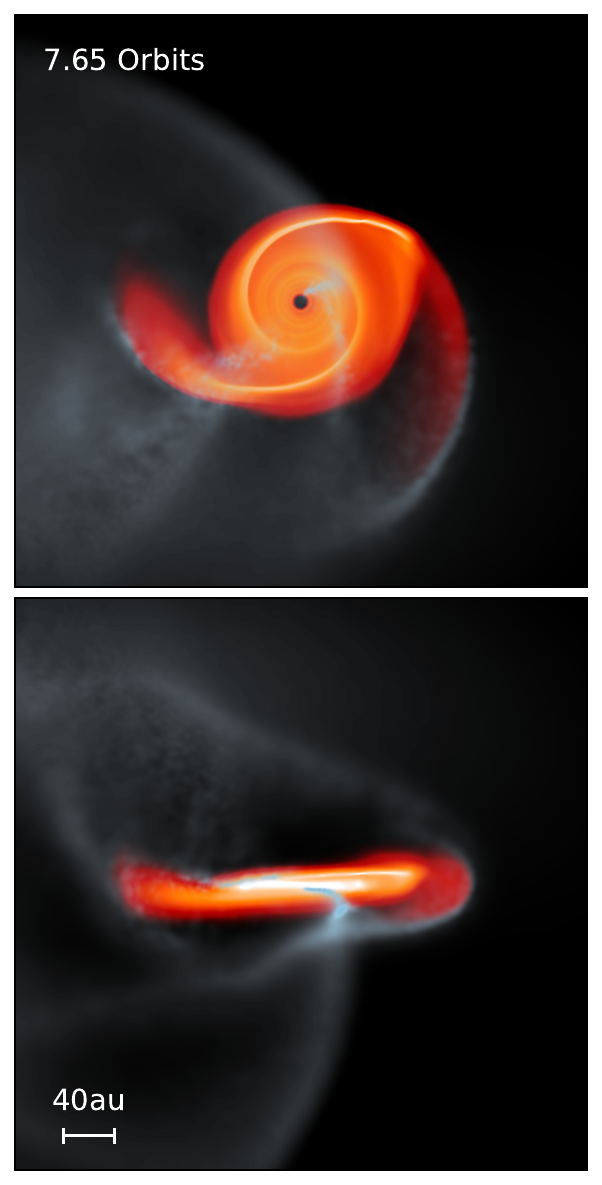}
\captionof{figure}{The particles in the simulation are classified as disc or cloud particles. The orange and blue rendered regions show the surface density calculate using just the disc and cloud particles, respectively.}
\label{fig:disc_v_cloud}
\vspace{1em}
\end{center}
Particles are considered to be part of the disc if its azimuthal velocity is near Keplerian as described by Eq. \ref{eq:disc}. The regions rendered in orange in Figure \ref{fig:disc_v_cloud} show the particles that satisfy this criterion and are labelled as disc particles. The remaining particles shown by the regions rendered in blue are considered to part of the infalling cloud.




\bsp	
\label{lastpage}
\end{document}